\documentstyle[12pt]{article}

\input{tcilatex}

\begin{document}

\title{On a time varying fine structure ''constant`` }
\author{Marcelo S.Berman$^{(1)}$ \and and Luis A.Trevisan$^{(2)}$ \\
(1) Instituto de Tecnologia do Paran\'{a}-- Tecpar\\
Grupo de Projetos Especiais\\
R.Prof Algacyr M. Mader 3775--CIC-CEP 81350-010\\
Curitiba-PR-Brazil\\
Email: marsambe@tecpar.br\\
(2) Universidade Estadual de Ponta Grossa,\\
Demat, CEP 84010-330, Ponta Grossa,Pr,\\
Brazil \ email: latrevis@uepg.br}
\maketitle

\begin{abstract}
Webb et al.%
\'{}%
s result that the fine structure 
\'{}%
\'{}%
constant'' $\alpha $ varies with time, is here considered due to
time-varying electric permittivity $\varepsilon _{0}$, along with an
inverselly varying magnetic permeability $\mu _{0},$ so as to keep the speed
of light $(1/\sqrt{\varepsilon _{0}\mu _{0}})$ constant. With help of Dirac%
\'{}%
s LNH, we find how the total number of nucleons of the Universe, the energy
density and Newton%
\'{}%
s Gravitational''constant%
\'{}%
\'{}%
evolve with time. We also estimate the present day value of the Universe'
deceleration parameter finding a value compatible with the Supernovae
observations, and we also found an acceptable time variation for the the
cosmological ''constant''.

PACS 98.80 Hw.

\newpage
\end{abstract}

\begin{center}
\bigskip ON \ A \ TIME \ VARYING \ FINE-STRUCTURE ``CONSTANT''

MARCELO S. BERMAN and \ LUIS A. TREVISAN

\bigskip
\end{center}

Webb et al. \cite{1} and Webb et al. \cite{2} have provided experimental
data on quasars that span \ 23\% to 87\% \ of the age of Universe, finding \
deviation from the average in  the fine structure constant, given by $\frac{%
\Delta \alpha }{\alpha }\simeq -0.72\times 10^{-5}.$ Due the fact that this
\ \ ``constant '' $(\alpha )$ is defined by other ones, one can ask what is
the ``constant '' that is causing the variation in $\alpha .$ Another
interesting remark is that this discovery is relating micro and macro \
phenomena and can have implications in the comprehension of QFT and
Cosmology. There are a variety of possible physical expressions for a
changing $\alpha .$ Bekenstein proposed a varying $e$ theory \cite{3}. An
alternative is the varying speed of light (VSL) theory \cite{4} in which
varying $\alpha $ is expressed as a variation of the speed of light.

Berman and Trevisan \cite{5} elaborated a full model containing a JBD
(Jordan-Brans-Dicke) framework with \ time-varying speed of light. Berman
and Trevisan \cite{6} returned to the subject commenting that similar
conclusions could be reached by applying Dirac%
\'{}%
s LNH (Large Number Hypothesis) with $c=c(t)$. We now present a different
scenario with $\alpha $ variable and LNH, \ keeping the speed of light \ and
Planck's constant really constants. One could claim that it is needed a
specific gravitational theory in order to deal with this project; however,
it is not certain which theory is the correct one for explaining
gravitational phenomena, so we feel that Dirac%
\'{}%
s hypothesis can guide us tentatively in the absence of a final theory. For
a full appraisal of LNH we recomend Barrow%
\'{}%
s article \cite{24}. Even a famous  researcher as Richard P. Feynman \cite
{25} admitted that among the speculators in numerical coincidences, there
were \ \ ''very serious mathematical players who construct mathematical
cosmological models ''. \ 

In S.I. units, the fine-structure ``constant'' , $\alpha $ is given by:

\begin{equation}
\alpha \equiv \frac{e^{2}}{2\varepsilon _{0}hc}
\end{equation}
We shall consider a possible time variation of constant $\varepsilon _{0}$
.If overdots stand for time derivatives, \ we have

\begin{equation}
\frac{\dot\alpha }{\alpha }=-\frac{\dot\varepsilon _{0}}{\varepsilon _{0}}
\end{equation}

Let us suppose that $\varepsilon _{0}$ varies with a power law of time,

\begin{equation}
\varepsilon _{0}=At^{n}
\end{equation}
($A,n=consts).$ Then ,

\begin{equation}
\frac{\dot\alpha }{\alpha }=-nt^{-1}
\end{equation}

The present Universe has been thought as an Einstein-de Sitter, with
constant deceleration parameter $q=1/2.$ We may ask whether the value of
this parameter could be a different constant value, say,

\begin{equation}
q=-\frac{\ddot{R}R}{\dot{R}^{2}}=m-1
\end{equation}
where $m$ is a constant to be determined and $R=R(t)$ is the scale factor in
Robertson-Walker%
\'{}%
s metric.

The theory of constant q's has been developed by Berman\cite{7}, and Berman
and Gomide\cite{8}, who found that the age of Universe, $t,$Hubble%
\'{}%
s parameter $H=\dot{R}/R,$ and constant $m$, are related by:

\begin{equation}
H=(mt)^{-1}
\end{equation}

It \ should be remarked that this formula independs on the particular
gravitational theory being \ considered. It is a property valid for
Robertson-Walker%
\'{}%
s metric, and it is approximately valid also for slowly time varying
deceleration parameters. \ 

The experimental value found by Webb et al, may be interpreted as yielding
the following result :

\begin{equation}
\frac{\Delta \alpha }{\alpha \Delta t}\simeq -\frac{0.72\times 10^{-5}}{0.64t%
}\simeq -1.1\times 10^{-5}Hm
\end{equation}
Notice that, in the above formula $\Delta t$ =$0.87t-0.23t=0.64t$.

Even if our numerical estimate as above turns out to be incorrect, we shall
\ employ it with the cautionary note that if a different numerical value
will \ be published later, any competent reader will be able to remake our
calculations, thus obtaining more accurate results than ours below.  

From (4) and (7), we find

\begin{equation}
n\simeq 1.1\times10^{-5}
\end{equation}
We have, thus, found how $\varepsilon _{0}$ must vary in order to comply
with experimental data \ on $\dot\alpha ,$ provided that all other constants
that appear in (1) are really constants.

From electromagnetism, we know that:

\begin{equation}
c=\frac{1}{\sqrt{\varepsilon _{0}\mu _{0}}}
\end{equation}
so that we find that $\mu _{0}$ should vary inversely as $\varepsilon _{0},$
in order to keep $c$ constant.

We might now take a look at Dirac%
\'{}%
s \ large number hypothesis \cite{9}\cite{10}\cite{11}, to check how of our
results could be accommodated in his framework. Calling $N$ the total number
of nucleons in the Universe, we have ,

\begin{equation}
\frac{cH^{-1}}{4\pi \varepsilon _{0}\left( \frac{e^{2}}{m_{e}c^{2}}\right) }%
\cong \sqrt{N}
\end{equation}
where \ $m_{e},m_{p}$ and $e$ stand respectively for electron 
\'{}%
s and proton%
\'{}%
s masses, and electron 
\'{}%
s charge,

\begin{equation}
\frac{e^{2}}{4\pi \varepsilon _{0}Gm_{p}m_{e}}\cong \sqrt{N}
\end{equation}
and

\begin{equation}
\frac{\rho (cH^{-1})^{3}}{m_{p}}\cong N
\end{equation}
where $\rho $ is the energy density of the Universe. The present value of $N$
is roughly 10$^{80}.$

Eddington \cite{Edd}proposed to consider another large number involving the
cosmological ''constant'', i.e,

\begin{equation}
ch(m_{n}m_{e}/\Lambda )^{1/2}\cong \sqrt{N}
\end{equation}
Then, on considering that $N$ increases with the age of the Universe, we
find that $\Lambda $ is time-varying. The whole hypothesis was coined by
Berman \cite{Berman} as GLNH (Generalized Large Numbers Hypothesis).

By plugging relations (6), (3) and taken care of result (8), we can check
that:

\begin{equation}
N=At^{1.99998}
\end{equation}
where $A$ is a constant. 
\begin{equation}
G=Bt^{-1.0}
\end{equation}
where $B$ is a constant. 
\begin{equation}
\rho =Dt^{-1.00002}
\end{equation}
where $D$ is a constant. 
\begin{equation}
\Lambda =Et^{-1.99998}
\end{equation}
where $E$ is a constant. Next generation of experimentalists may well
provide evidence in favor or against these results. The time variation
obtained for $\Lambda ,$is compatible with our knowledge about the value
that it should have at GUT%
\'{}%
s time and in the present.

It is necessary to point out that the origin of $c=c(t)$ theories can be
traced to a paper by Gomide \cite{12}, and that $\dot{\alpha}\neq 0$ theory
with $c=c(t)$ was considered by Barrow and Magueijo \cite{13}. Gomide also
worked with a time varying $\varepsilon _{0},$ but supposed that $\alpha $
was constant, in face of Bahcall and Schmidt%
\'{}%
s paper \cite{14}.

\bigskip Confronting with observations, i.e., when we define

\begin{equation}
\frac{\dot G}{G}=\sigma H
\end{equation}
our result is, from (14),

\begin{equation}
\frac{\dot G}{G}\cong -1.0t^{-1}=-1.0H(1+q)
\end{equation}
where we have, again, used relations (6) and (14).

This means that, if we would have accurate measures of $\sigma $, we could
estimate the deceleration parameter $q.$ However, we refer to Will \cite{15} 
\cite{16}, in order to mention that there is no conclusive experimental
value for $\sigma .$ Lunar laser ranging and Viking radar measurements by
Williams et al\cite{17}and Reasenberg\cite{18} put \TEXTsymbol{\vert}$\sigma
|$ $<0.6$. In Ref.\cite{16}, Will coments that these two kinds of
measurements give the best limits on $\dot G/G$ \cite{20}.

This means, \ from (17), that:

\begin{equation}
-0.4>q>-1.6
\end{equation}
so that the Universe would be accelerating, in accordance with Supernovae
results \cite{19}. We have thus shown that Dirac%
\'{}%
s LNH, Webb et al%
\'{}%
s fine structure constant time variation, Supernovae results and $\dot{%
\varepsilon}_{0}$ $\neq 0$ hypothesis are all coherent among them. It could
be argued that there is one evidence \cite{21} for a non constant $q$
arising from a ten billion years old Supernova explosion; however, this is
not a conclusive evidence for turning down the constant $q$ hypothesis for
the present Universe. Just as we have discussed above the time variation for 
$G,$ from the experimental point of view, we might comment on the time
variation of $\rho $ as found by us. Unfortunately, it is very difficult to
estimate with accuracy the average density of the present Universe, and then
we are left without experimental clues on $\dot{\rho}$ .Nevertheless, we
have found a time decreasing function for $\rho $ ; this is the kind of
variation unanimously expected by researchers in the field, i.e, no one
would favour an increasing function of time, because as the Universe
expands, $\rho $ should decrease.

\ We have thus found how $N,G,\rho ,\Lambda $ and $\varepsilon _{0}$ may
vary and we have found bounds on the \ present day deceleration parameter $q$
in agreement with observation. In the model presented here, we have $\alpha
=\alpha (t),$ because $\varepsilon _{0}=\varepsilon _{0}(t),$ but $c$ is
constant. It's important to elaborate different models and make some
previews about their consequences in order to decide among them. The way,
and why, $\alpha $ is varying with the age of Universe is one of most
intrigating problem in \ modern physics and the understanding of this
question can lead to new discoveries. In fact, a Superunification theory
will only survive in case that such variations of constants with the age of
the Universe, shall encounter with a theoretical explanation. 

It is necessary to point out that we did not endeavour to make precise
numerical predictions on the values of the quantities we did estimate,
because GLNH is not a substitute for an exact gravitational theory. In fact,
we only used GLNH for obtaining tentative  laws of variation for these
quantities, with the age of the Universe. 

{\bf Acknowledgements}

The authors are recognized to an anonymous source, for clarifying our
manuscript . We also thank Prof. F.M.Gomide for supplying pertinent
information and Prof M.M. Som for his long time support.

Both authors thank support by Prof. Ramiro Wahrhaftig, Secretary of Science,
Technology , and Higher Education of the State of Paran\'{a}, and by our
Institutions, especially to Jorge\ L.Valgas, Roberto Merhy, Mauro K.
Nagashima, Carlos Fior, C.R. Kloss, J.L.Buso, and Roberto Almeida.

\bigskip

\end{document}